# Theoretical study of charge exchange dynamics in $He^+$ + NO collisions

**E. Bene[1] and M.-C. Bacchus-Montabonel[2]**

*1. Institute for Nuclear Research, Hungarian Academy of Sciences (MTA Atomki), P.O. Box 51, H-4001 Debrecen, Hungary*
*2. Institut Lumière Matière, UMR5306 Université Lyon 1-CNRS, Université de Lyon 69622 Villeurbanne Cedex, France*

**ABSTRACT**

We investigate the charge transfer mechanism in the collisions of helium ions on nitric oxide using a molecular description framework with consideration of the orientation of the projectile toward the target. The anisotropy of the collision process has been analysed in detail in connection with the non-adiabatic interactions around avoided crossings. Potential energy curves, radial and rotational coupling matrix elements have been determined by means of *ab initio* quantum chemical methods. The collision dynamics is performed in the [1.-25.] keV collision energy range using a semiclassical approach, and the total electron transfer cross sections are analysed with regard to available experimental data.

## I. INTRODUCTION

Studies on collisions of diatomic molecules like $N_2$, $O_2$, CO and NO with light hydrogen and helium ions are of fundamental importance for understanding the processes in natural and man-made plasmas, and for modelling planetary and cometary atmospheres (see e.g. Refs. [1-6]). The action of proton and He ion impact on these molecules has been widely investigated experimentally, mainly in the energy region below 100 keV where electron capture has been shown to be an important reaction, complementary to the excitation, ionization and fragmentation processes [7-10]. Siegmann et al. [11] measured the relative cross sections for multiple ionizations in collisions of 50−300 keV protons, deuterons and helium ions with $N_2$, $O_2$, CO and NO molecules, and studied the angular distributions of molecular fragments for each ionization channel. Furthermore, they presented theoretical calculations within the statistical deposition model [12] with a conclusion that, at small velocities, the experimental and theoretical results disagree. Such failure may be explained by introduction of the electron capture mechanism. However, due to missing data in the literature on the capture rates for separate ionization channels in these collisions, it is very important to give any estimate for charge transfer processes either experimentally or theoretically.

In this paper, our goal is to undertake a theoretical study of such processes in the keV energy range. We have concentrated on the charge exchange dynamics in the collisions of $He^+$ ions with nitric oxide molecules in the framework of the molecular representation of the collisions. Such collisions are interesting from a biophysical point of view. Considering that NO plays an important role in a variety of biological processes, collisions with $He^+$ ions are relevant in the radiation damage of living tissues and in the irradiation of the biological medium [13,14].

There are a couple of experimental measurements available in the literature on the $He^+ - NO$ collisions. Most of the materials are devoted to the charge-exchange process. In the medium-energy range, total charge transfer cross sections for neutralization of $He^+$ ion beams were measured for 1. − 3. keV ions by Moran and Conrads [15]. They used time-of-flight methods to identify the forward scattered $He^0$ products from the $He^+$ interactions with several molecular targets, and, for the $He^+ - NO$ collisions, they measured large total electron capture cross sections suggesting that a near resonant process occurs. The $NO^+$ excited states involved in the charge transfer process could not be identified, however, from the measurement of the relative abundance of the $N^+$, $O^+$, and $NO^+$ products. It was inferred that the molecular ions were formed dominantly in dissociative or predissociation states [16]. Barat and co-workers [17], also by means of a time-of-flight technique, presented energy-loss measurements for the collisions of the He and $He^+$ projectiles with CO and NO targets in the 0.5–3. keV energy range, and relative differential cross sections were determined in the scattering angle range $0^\circ - 3^\circ$. The energy-loss measurements locate the relevant states of the molecular ion in the Franck-Condon region, allowing us to obtain information on the reaction channels. The paper of Gao *et al*. [18] reported measurements of absolute differential cross sections for charge transfer in collisions of $H^+$ and $He^+$ with several molecules over the laboratory angular range $0.02^\circ - 1.0^\circ$. The cross sections were determined at 1.5 keV for collisions of $He^+$ with $H_2$, $N_2$, $O_2$, CO and NO.

From a theoretical point of view, a quantum dynamics study of the charge-transfer in the $He^+$ + NO collision is still lacking. Recently, Amaran and Kumar [19] carried out a time-independent quantum dynamics study of the $H^+$ + NO collision considering the vibrational charge transfer processes at small collision energies, 9.5 and 29.03 eV, within the vibrational close-coupling rotational infinite order sudden approximation (VCC-RIOSA) approach using the coupled ground state and first excited state *ab initio* quasidiabatic PES's of Amaran et al. [20]. In the present study we have applied a procedure similar to those employed in Refs. [21-24] for collisions of $C^{2+}$ ions with the diatomic targets OH, CO, HF and HCl; and recently by Rozsályi *et al.* [25] for $H^+$ + $N_2$ collisions. The charge exchange process is analysed by the use of a molecular description model which considers the evolution of a projectile-target quasimolecule and provides information on the mechanism and on the electronic structure of the projectile and target during the reaction. High-level *ab initio* molecular computations of the potential energy curves of the $HeNO^+$ quasimolecule and the corresponding nonadiabatic couplings are performed augmented with semiclassical dynamical calculations in the [1.-25.] keV collision energy range. The mechanism of such a process depends of course significantly on the orientation of the target toward the projectile. In several previous studies, electron transfer has been shown to be highly anisotropic with electron delocalization from the target to the colliding ion [21-28]. We have thus focused our attention on the anisotropy of the $He^+$-NO collision system by looking at a set of fixed orientations. Such a study cannot lead to a direct comparison with experiment as measured cross sections correspond to an average on all possible orientations. However, the analysis of a series of given geometries with regard to available experimental data [15] may drive some conclusions on the charge transfer mechanism.

This article is organized as follows. In section II we briefly overview the basic equations of the theoretical method. In Sec. III we give the computational details on the molecular calculations and potential energy curves are presented and discussed through avoided crossings. The charge transfer cross sections are reported with regard to the experimental total cross sections of Ref. [15] in Sec. IV. Finally, we summarize our numerical results along with a conclusion.

## II THEORETICAL TREATMENT

The $HeNO^+$ molecular system can be described by the internal Jacobi coordinates $\{\mathbf{R},\boldsymbol{\rho},\theta\}$, where $\mathbf{R}$ is the distance of $He^+$ from the center of mass of the nitric oxide molecule, $\boldsymbol{\rho}$ is the interatomic distance of NO, and $\cos\theta = \hat{\rho}\hat{R}$ (see Fig. 1).

In the impact energy range we are dealing with, semiclassical approaches may be used with a good accuracy to describe the ion-molecule collision dynamics. The electronic motion is thus described in a quantum mechanical treatment, while the nuclei follow classical trajectories. As classical straight-line trajectories are satisfying for energies greater than 10 eV/amu [29], we applied the impact parameter or eikonal approximation, where $\mathbf{R} = \mathbf{b} + \mathbf{v}t$ with impact parameter **b** and constant velocity **v**. The remaining degrees of freedom are treated by the eikonal equation:

$$\left( H_{\text{int}} - i\frac{\partial}{\partial t}\bigg|_{\mathbf{r},\boldsymbol{\rho}} \right)\Psi(\mathbf{r},\boldsymbol{\rho},t) = 0 \qquad (1)$$

where $\mathbf{r}$ stands for the electronic coordinates, and $H_{\text{int}}(\mathbf{r},\boldsymbol{\rho};\mathbf{R})$ denotes the internal Hamiltonian obtained by subtracting from the total Hamiltonian the nuclear kinetic energy term associated with $\mathbf{R}$.

The collision dynamics has been treated in the framework of the sudden approximation hypothesis considering electronic transitions much faster than rotation and vibration motions. The molecular target may thus be assumed to be frozen during the collision time and the ion-molecule collision system may be visualized as an ion bumping into an anisotropic atom. The total and partial cross sections, corresponding to purely electronic transitions are thus determined by solving the impact-parameter equation as in the usual ion-atom approach with a fixed internuclear distance of the target molecule. This first order approximation has been widely used and has shown its efficiency in the field of ion-diatomics [30-32], and even in ion-polyatomics collisions [26-28] in the keV energy range.

In the eikonal-sudden approach (see [33]) we can substitute $H_{int}$ by the clamped-nuclei electronic Born-Oppenheimer Hamiltonian $H_{el}$ and the wave function of the collisional system can be expressed as [34]:

$$\Psi_{LM} = \rho^{-1} Y_{LM}(\hat{\rho})\chi_0(\rho)\psi(\mathbf{r};\boldsymbol{\rho},t) \qquad (2)$$

where $Y_{LM}$ are the spherical harmonic functions, $\chi_0$ is the initial vibrational wave function, and $\psi(\mathbf{r};\ \boldsymbol{\rho},\ t)$ is the electronic wave function, which can be expanded in terms of the eigenfunctions $\{\varphi_i\}$ of $H_{el}$ with the eigenvalues $\{\varepsilon_i\}$:

$$\psi = \sum_j a_j(t;\boldsymbol{\rho})\varphi_j \exp\left[-i\int_0^t \varepsilon_j dt'\right], \qquad (3)$$

where

$$H_{el}\varphi_j(\mathbf{r};\boldsymbol{\rho},\mathbf{R}) = \varepsilon_j(\mathbf{R},\boldsymbol{\rho})\varphi_j(\mathbf{r};\boldsymbol{\rho},\mathbf{R}). \qquad (4)$$

For a given nuclear trajectory and fixed $\boldsymbol{\rho}$, substitution of (3) and (2) into (1), and integration of the equation leads to the following system of coupled differential equations:

$$i\frac{da_j}{dt}=\sum_k a_k \left\langle \varphi_j \left| H_{el} - i\frac{\partial}{\partial t} \right| \varphi_k \right\rangle \exp\left[ -i\int_0^t (\varepsilon_k - \varepsilon_j) dt' \right], \qquad (5)$$

where the values of the integrals $\langle \varphi_j | \partial/\partial t | \varphi_k \rangle$ are the dynamical (or non-adiabatic) coupling terms.

Using the trajectory relation, with the Z-axis of the laboratory reference frame pointing along the direction of **v**, we can obtain a two-component expression for the dynamical coupling terms:

$$\langle \varphi_j | \frac{\partial}{\partial t} | \varphi_k \rangle = \frac{vR_z}{R} \langle \varphi_j | \frac{\partial}{\partial R} | \varphi_k \rangle - \frac{vb}{\mathrm{R}^2} \langle \varphi_j | iL_y | \varphi_k \rangle, \quad (6)$$

where $\langle \varphi_j | \partial/\partial R | \varphi_k \rangle$ and $\langle \varphi_j | iL_y | \varphi_k \rangle$ are the radial, and rotational coupling matrix elements, respectively. This leads to a set of coupled differential equations:

$$i\frac{da_j(t)}{dt}=\sum_k a_k(t)\left( \langle \varphi_j | H^{el} | \varphi_k \rangle - i\frac{vZ}{R} \langle \varphi_j | \frac{\partial}{\partial R} | \varphi_k \rangle - i\frac{vb}{R^2} \langle \varphi_j | iLy | \varphi_k \rangle \right) \exp\left( -i\int_0^t (\varepsilon_k - \varepsilon_j) dt' \right) (7)$$

In the dynamical calculations an approximation is employed, that is we fix the molecular orientation during the trajectory [25,33]. Namely, we assume that the couplings and the energies $\varepsilon_j$ only depend on the internal nuclear coordinate *R*, while $\theta$ and $\rho$ are kept as fixed parameters. For the $\{ a_j(t;\rho,\theta,b,v) \}$ coefficients, this assumption leads to a system of differential equations similar to that obtained for ion-atom collisions.

The non-adiabatic coupling terms can be obtained in electronic structure calculations together with the potential energy surfaces $\varepsilon_j(R;\rho,\theta)$, and they are then substituted into (6) to obtain the expansion coefficients $a_j(t;\rho,\theta,b,v)$ for each velocity and impact parameter.

The isotropic probability on each capture channel $f$ may be given by $P_f(b,v;\rho,\theta) = \lim_{t\to\infty} |a_f(t;\rho,\theta,b,v)|^2$, and the partial isotropic cross section may be defined as

$$\sigma_f(v,\rho,\theta) = 2\pi \int_0^\infty P_f(b,v;\rho,\theta) b db. \quad (8)$$

The total cross section is then obtained by $\sigma_{tot} = \sum_f \sigma_f$ with summation over all charge exchange channels.

Note that the present form of the isotropic cross section (8) is derived for the case of homonuclear targets [33], however, this formula can be used here as well due to the fact that the atomic masses of our target atoms are almost equal.

With regard to the limited basis used in the calculation, the set of $\{\varphi_i\}$ should be modified by appropriate translation factors [35] to avoid spurious coupling terms at long range and to assume cross sections to be independent of the origin of coordinates. Such effect may be evaluated in the approximation of the common translation factors [36, 37] where the radial and rotational coupling matrix elements may be transformed, respectively, into:

$$\langle \varphi_j \left| \partial / \partial R - (\varepsilon_j - \varepsilon_k) z^2 / 2R \right| \varphi_k \rangle \quad (9)$$

and

$$\left\langle \varphi_j \left| iL_y + (\varepsilon_j - \varepsilon_k) zx \right| \varphi_k \right\rangle, \quad (10)$$

where $z^2$ and zx are the components of the quadrupole moment tensor.

## III. MOLECULAR CALCULATIONS

The potential energy curves and the non-adiabatic coupling matrix elements (NACME) were computed as functions of the variable *R*, whereas *ρ* was kept fixed at its optimized value in the ground state. In order to study the anisotropy of the charge transfer mechanism, given geometries of the $He^+$-NO collisional system have been investigated : the two collinear orientations, with $He^+$ approaching either the oxygen ($\theta$=0°) or the nitrogen atom ($\theta$=180°), together with non-linear orientations, the perpendicular approach ($\theta$=90°) and the geometry corresponding to the angle $\theta$=45°.

As spin orbit coupling is negligible in the energy range of interest, we assumed that electron spin was conserved in the collision process. The NO ground state being of $^2\Pi$ symmetry, the $He^+(1s^1)^2S$ + $NO(^2\Pi)$ entrance channel corresponds to $^1\Pi$ and $^3\Pi$ states of the linear $HeNO^+$ quasimolecule, so singlet and triplet states may be correlated with the entrance channel. The computations have been performed for the singlet and triplet spin symmetries in the $C_{2v}$ point group for the collinear approaches and in the $C_s$ group for non-linear approaches. In the latter case, the plane of the molecular system was considered to be the plane of symmetry. The molecular calculations have been performed by means of the quantum chemistry software package MOLPRO [38]. For He, N, and O atoms we employed the extended Dunning's correlation consistent polarized valence triple-ζ basis set (*aug*-*cc*-*p*VTZ) [39]. The natural orbitals were optimized at the state-averaged Complete Active Space Self-Consistent Field (CASSCF) level of theory. The nitrogen and oxygen 1*s* orbitals being doubly occupied, the remaining twelve valence electrons were distributed on 13 orbitals, thus in the $C_{2v}$ point group, the active space included two $a_1$ inactive orbitals, and seven $a_1$, three $b_1$, and three $b_2$ active orbitals. For both spin symmetries, the molecular states were treated with the same weight for each $A_1$, $B_1$, and $A_2$ symmetry. Correspondingly, in the $C_s$ point group the lowest two $a'$ orbitals being kept closed, the $3a'$–$12a'$ and $1a''$–$3a''$ orbitals were active.

The charge transfer process is driven mainly by nonadiabatic interactions in the vicinity of avoided crossings(see e.g. [40]). We have thus to take into account all the exit channels corresponding simultaneously to the excited states of the $He^+$ ion and also all the possible excited states of $NO^+$, which could be correlated with the entrance channel by means of radial or rotational coupling. In all cases, the radial coupling matrix elements between all pairs of states of the same symmetry have been calculated by means of the finite difference technique:

$$\left\langle \varphi_j \left| \partial / \partial R \right| \varphi_k \right\rangle = \left\langle \varphi_j(R) \right| \lim_{\Delta \to 0} \frac{1}{\Delta} \left| \varphi_k(R+\Delta) - \varphi_k(R) \right\rangle, \quad (11)$$

which, taking account of the orthogonality of the eigenfunctions $\left| \varphi_j(R) \right\rangle$ and $\left| \varphi_k(R) \right\rangle$ for $j \neq k$ reduces to

$$\left\langle \varphi_j \left| \partial / \partial R \right| \varphi_k \right\rangle = \lim_{\Delta \to 0} \frac{1}{\Delta} \left\langle \varphi_j(R) \middle| \varphi_k(R+\Delta) \right\rangle, \quad (12)$$

with the previously tested parameter Δ=0.0012 a.u. [41] and using the three-point numerical differentiation method for sake of good numerical accuracy.

The rotational coupling matrix elements $\langle \varphi_j(R) | iL_y | \varphi_k(R) \rangle$ between states of different space symmetries have been calculated directly from the quadrupole moment tensor using the expression $iL_y = x\frac{\partial}{\partial z} - z\frac{\partial}{\partial x}$ by taking the centre of mass of the system as the origin of electronic coordinates [42].

In the linear cases, the $He^+(1s^1)^2S$ + $NO(^2\Pi)$ entrance channel corresponds to $^1\Pi$ and $^3\Pi$ states, $^{1,3}\Pi$ states can thus be correlated to the entrance states by means of radial couplings, and $^{1,3}\Sigma$ and $^{1,\,3}\Delta$ states can be correlated by means of rotational coupling interaction. With regard to the respective ionization potentials and different excited states of He and $NO^+$, a number of states may be considered, with only the $^1S$ ground state of helium involved in the charge transfer process. Tables 1 and 2 show the different channels we have to take into account for the singlet, and triplet spin symmetries, respectively. The asymptotic energies for $HeNO^+$ (linear geometry, $\theta$=180°) at 12 a.u. compare well with the CASSCF and MRCI calculations of separated species. We used experimental data for the helium ion [43], and the geometry optimization for the target molecule was performed with the MultiReference Configuration Interaction (MRCI) method. The optimized geometry of the $^2\Pi$ ground state of NO is $r_{NO}$=2.187 a.u., in concordance with the experimental value 2.175 a.u. [44] and corresponding to a vertical ionization potential of 9.28 eV in good agreement with the experimental value of 9.26 eV [44].

The potential energy curves of the $HeNO^+$ system and the associated radial and rotational couplings were calculated in the 2-12 a.u. internuclear distance range. The potential energy curves for the $(He – NO)^+$ system corresponding to a linear geometry with the $He^+$ ion approaching the nitrogen atom ($\theta$=180°) are presented in Fig. 2a,b. With regard to the different excited states of $NO^+$, nine electronic states of the linear $HeNO^+$ molecular system must be taken into account in the charge transfer process for the singlet multiplicity (see Table 1). Four $^1\Pi$ states, including the entrance channel $^1\Pi\{He^+ + NO(^2\Pi)\}$, can be correlated by means of radial coupling. Furthermore, there are two $^1\Sigma^+$, one $^1\Sigma^-$, and two $^1\Delta$ states correlating with the entrance channel by rotational coupling interaction. For triplet spin symmetry, ten states must be considered: five $^3\Pi$ states (with the entrance channel $^3\Pi\{He^+ + NO(^2\Pi)\}$, one $^3\Sigma^+$, two $^3\Sigma^-$ and two $^3\Delta$ levels (see Table 2). In both spin multiplicities, this leads to a very intricate molecular system involving a great number of molecular states with numerous avoided crossings in short internuclear distance range, below R=5 a.u.. The charge transfer involving a singly charged projectile ion induces indeed very short range interactions, some of them in the repulsive part of the potential energy curves. Very sharp avoided crossings corresponding to narrow energy gaps are exhibited between the $^1\Pi$ entrance channel and the higher $^1\Pi\{He + NO^+(3^1\Pi)\}$ charge transfer level around R=4 a.u. for singlets, and between the $^3\Pi$ entrance channel and the $^3\Pi\{He + NO^+(4^3\Pi)\}$ exit channel for triplets, around 4.6 a.u.. Avoided crossings may also be observed between $^{1,3}\Pi$ states in the repulsive part of the potential energy curves, in particular for triplet states around R=3.4 a.u. between the entrance channel and the $^3\Pi\{He + NO^+(3^3\Pi)\}$ state. The process appears clearly mainly driven by radial couplings between Π states, either in the singlet or triplet manifold. However significant rotational coupling may occur between the He + $NO^+(1^1\Pi)$ level and both the He + $NO^+(1^1\Delta)$ and He + $NO^+(1^1\Sigma^-)$ states which correspond to the same asymptotic energy. Besides, smooth avoided crossings may be observed between the $\Sigma^+$ and Δ states, in particular between the $1^{1,3}\Delta$ and the $2^{1,3}\Delta$ around 3.5 a.u. in both triplet and singlet manifolds. The present calculation in the linear geometry toward nitrogen has been compared to the collision toward the oxygen atom and calculations have been performed for singlet and triplet spin symmetries for the collision of a $He^+$ ion approaching the oxygen atom ($\theta$=0°). The same features may be observed as already pointed out for the linear collision approaching the nitrogen atom ($\theta$=180°), however a stronger interaction between both He + $NO^+(^1\Delta)$ states, He + $NO^+(1^1\Delta)$ and He + $NO^+(2^1\Delta)$, is clearly exhibited around R=3.4 a.u. as shown in Fig. 3.

The potential energy curves have been also calculated for singlet and triplet manifolds for non-linear geometries, corresponding to a perpendicular approach ($\theta$=90°), and to an intermediate approach ($\theta$=45°). As pointed out previously, the $HeNO^+$ molecular system is extremely complex, involving a great number of molecular states with very short-range interactions, and extensive analysis of a great number of different geometries would be very heavy, we choose thus to focus on given characteristic orientations. In that case, we computed the radial and rotational couplings between eight A′ and seven A″ states for singlets, and between eight A′ and nine A″ levels for triplets in the $C_s$ symmetry group. Interesting short range avoided crossings may be pointed out between $^{1,3}A'$ states, correlated either to $^{1,3}\Sigma$ or $^{1,3}\Pi$ $NO^+$ configurations. They are presented in Fig.4a for the singlet manifold in perpendicular geometry. Such avoided crossings around R=3 a.u. appear shifted towards shorter distances with regard to the interactions observed in the linear geometries. The same feature may also be observed for the triplet manifold. More generally, as shown on Fig.4 a,b, a regular shift of such avoided crossing is observed from linear, toward the angle $\theta$=45° and the perpendicular geometry, in both singlet and triplet spin manifolds.

## IV. DYNAMICAL CALCULATIONS

The collision treatment has been performed in the [1.−25.] keV laboratory energy range ([0.1−0.5] a.u. collision velocities) in the eikonal-sudden approximation approach. The partial and total cross sections averaged over the impact parameter have been calculated between the different quasimolecular states involved in the charge transfer process using the EIKONXS code based on an efficient propagation method [45]. The calculations have been performed separately for singlet and triplet manifolds taking into account all the radial and rotational couplings. The total cross section is then calculated with regard to the statistical weights ¼ and ¾ for singlet and triplet spin symmetries respectively by the expression $\sigma_{tot} = \frac{1}{4}\,{}^{1}\sigma + \frac{3}{4}\,{}^{3}\sigma$, ${}^{1,3}\sigma$ corresponding to the cross sections for the singlet and triplet manifolds respectively. Translation factors have not been included in this study. However, the chosen origin coordinate may be expected to provide accurate enough values of total cross sections for impact velocities lower than 0.5 a.u. as those we are investigating [46,47], even if transitions driven by rotational couplings could be slightly overestimated. Such translation effect has besides been tested in the linear approach ($\theta$=180°) by comparing the collision dynamics for $^3He^+$-NO and $^4He^+$-NO at the same velocity. The calculations show a discrepancy lower than 2%, even at collision energies up to 100 keV, assuming a very low effect for the $He^+$-NO collision system, in particular in the [1.-25.] keV collision energy range. Such result is also confirmed by calculations considering the translation effect by means of the common translation approximation [35]. The translation effect can thus be considered as negligible in our collision system, as already pointed out in a number of ion/atom [48] and ion/molecule [23] collisional systems and has not been taken into account for the other geometries, with regard to the great number of molecular states involved in the process.

The results of the dynamical calculations for the perpendicular geometry are presented in Fig.5 together with the experimental results of Moran and Conrads [15]. The charge transfer cross sections for both spin symmetries show a regular decrease with increasing collision energy. Although all orientations should be taken into account for a direct comparison with experimental data, our calculation at a given geometry may compare rather positively with the experimental features, reproducing even the two steps exhibited by the experimental cross section. The agreement with experiment is however not completely satisfactory and a shift in the cross section values is exhibited, which may be first of all relied to the fact that we are looking here to the cross section values calculated at a given geometry, but also to the uncertainties involved in our calculation with regard to the complex molecular system we are dealing with. Effectively, for a collision with a singly charged projectile ion as $He^+$-NO, the curve crossings take place at very short range distances where the energy gap at the crossing point is very sensitive, in particular in the repulsive part of the potential curve. We have also to consider the uncertainties on the experimental results, which can reach up to 20-30% usually in such systems [49]. With regard to the statistical weight, the mean cross section values are very close to the cross sections calculated for the triplet manifold. The charge transfer cross sections on the different space symmetries A' and A'' are detailed in Table 3. In both spin symmetries, the process is clearly driven by the non-adiabatic radial coupling interactions, the rotational effect remains low all over the collision energy range, even if it increases slightly at higher energies.

An analysis of the anisotropic effect may be developed by looking at the behaviour of the charge transfer cross sections from perpendicular to linear orientations. The corresponding total cross sections taking account of the singlet and triplet manifolds are presented in Fig. 6. A clear evolution from the perpendicular to the linear approach may be pointed out showing a better agreement with experimental cross sections in the perpendicular approach. The cross sections in both $\theta=0°$ ( $He^+$ - ON) and $\theta=180°$ ($He^+$ - NO) linear geometries show very similar variations with higher values in the approach toward the nitrogen atom. Of course this analysis remains qualitative as all possible orientations of the projectile $He^+$ ion toward the molecular target should have to be taken into account in order to drive a direct comparison with experiment. Convolution over geometric parameters would present a low weight for linear orientations with regard to non-linear ones. Anyway this tendency is worth to be pointed out with regard to the charge transfer mechanism. Besides, such behaviour may be compared to our earlier studies on the orientation effect in the collisions of $C^{2+}$ ions with OH, CO, and hydrogen halide molecules [21-24]. In these collisions the charge transfer was favoured in the linear approach with reaction of the $C^{2+}$ ion toward the most electronegative atom and the perpendicular approach was markedly less efficient. In the present case, even if charge transfer cross sections are lower in the perpendicular approach, they remain quite significant. This could be related directly to the features of the molecular collision systems. Effectively, in collisions with multiply charged ions, the main contribution to the charge exchange cross section comes from relatively long range crossings, whilst for single charged projectile ions, avoided crossings occur at much shorter distances as the entrance and charge transfer channels have almost the same asymptotic behaviour. As the charge transfer mechanism depends fundamentally on the involved specific non-adiabatic interactions, a different behaviour for $He^+$-NO would thus be expected. We may point out in particular the interesting short range interaction in the perpendicular orientation exhibited in Fig. 4a.

## V. CONCLUSIONS

We present in this work an ab-initio treatment of the charge transfer process in collisions of $He^+$ ions with the NO molecule. The mechanism has been analysed through the different molecular states involved in the process in the different space and spin symmetries. The system appears very complex, involving a great number of molecular states and very short range non-adiabatic interactions as a singly charged projectile ion is concerned. In particular intricate avoided crossings have to be pointed out leading to time-consuming determination of radial and rotational coupling terms. The charge transfer process appears highly anisotropic and cross sections in the perpendicular orientation may be compared positively to the experimental data of Moran and Conrads [15]. Such features, driven by the presence of short range crossings could certainly be extended to other singly charged/molecule collision systems and show a different behaviour than charge transfer processes induced by multiply charged projectile ions.

## ACKNOWLEDGEMENTS

Support from the Hungarian National Science Foundation OTKA (Project No. K109440) is gratefully acknowledged. We acknowledge support from the COST actions MP1002 Nano-IBCT, CM0805 Chemical Cosmos and CM1204 XLIC.

## References

[1] F. A. Gianturco, U. Gierz, J. P. Toennies, J. Phys. B 14, 667 (1981).

[2] S. Wyckoff, J. Theobald, Adv. Space. Res. 9, 157 (1989).

[3] P. Reinig, M. Zimmer, F. Linder, At. Plasma Mater. Interaction Data Fusion 2, 95 (1992).

[4] C. J. Latimer, Adv. At. At. Mol. Opt. Phys. 30, 105 (1993).

[5] E. Marsch, X.-Z. Ao, C.-Y. Tu, J. Geophys. Res. 109, A04102 (2004).

[6] J. M. Ajello, R. S. Mangina, D. J. Strickland, D. Dziczek, J. Geophys. Res. 116, A00K03 (2011).

[7] F. B. Yousif, B. G. Lindsay, C. J. Latimer, J. Phys. B 23, 495 (1990).

[8] M. B. Shah, H. B. Gilbody, J. Phys. B 23, 1491 (1990).

[9] H. O. Folkerts, R. Hoekstra, R. Morgenstern, Phys. Rev. Lett. 77, 3339 (1996).

[10] A. Lafosse, J. C. Houver, D. Dowek, J. Phys. B 34, 819 (2001).

[11] B. Siegmann, U. Werner, Z. Kaliman, Z. Roller-Lutz, N. M. Kabachnik, and H. O. Lutz, Phys. Rev. A 66, 052701 (2002).

[12] N. M. Kabachnik, V. N. Kondratyev, Z. Roller-Lutz, and H. O. Lutz, Phys. Rev. A 56, 2848 (1997); 57, 990 (1998).

[13] C. von Sonntag, The Chemical Basis for Radiation Biology, Taylor and Francis, London, 1987.

[14] B. D. Michael, P.D. O'Neill, Science **287**, 1603 (2000).

[15] T. F. Moran and R. J. Conrads, J. Chem. Phys. 58, 3793 (1973).

[16] J. E. Parker and R. Y. Haddad, Int. J. Mass Spectrom. Ion. Phys. 27, 403 (1978); 31, 103 (1979).

[17] D. Dowek, D. Dhuicq, and M. Barat, Phys. Rev. A 28, 2838 (1983).

[18] R. S. Gao, L. K. Johnson, C. L. Hakes, K. A. Smith, and R. F. Stebbings, Phys. Rev. A 41, 5929 (1990).

[19] S. Amaran and S. Kumar, J. Chem. Phys. 128, 124306 (2008).

[20] S. Amaran, S. Kumar, and H. Köppel, J. Chem. Phys. 128, 124305 (2008).

[21] E. Bene, Á. Vibók, G. J. Halász, M.- C. Bacchus-Montabonel, Chem. Phys. Lett. 455, 159 (2008).

[22] E. Bene, P. Martínez, G. J. Halász, Á. Vibók, and M.- C. Bacchus-Montabonel, Phys. Rev. A 80, 012711 (2009).

[23] E. Rozsályi, E. Bene, G. J. Halász, Á. Vibók, and M.- C. Bacchus-Montabonel, Phys. Rev. A 81, 062711 (2010).

[24] E. Rozsályi, E. Bene, G. J. Halász, Á. Vibók, and M.- C. Bacchus-Montabonel, Phys. Rev. A 83, 052713 (2011).

[25] E. Rozsályi, L. F. Errea, L. Méndez, and I. Rabadán, Phys. Rev. A 85, 042701 (2012).

[26] M.-C. Bacchus-Montabonel, M. Łabuda, Y. S. Tergiman, J. E. Sienkiewicz, Phys. Rev. A 72, 052706 (2005).

[27]M.-C. Bacchus-Montabonel, Y. S. Tergiman, D. Talbi, Phys. Rev. A 79, 012710 (2009).

[28] M.-C. Bacchus-Montabonel, Y. S. Tergiman, Comput. Theor. Chem. 990, 177 (2012).

[29] B. H. Bransden, M. R. C. McDowell, in *Charge Exchange and the Theory of Ion- Atom Collisions* (Clarendon Press, Oxford, 1992), p. 63,64.

[30] D. R. Bates, R. McCarroll, Proc. Roy. Soc. A 245, 175 (1958).

[31] M. Kimura, N. F. Lane, Adv. At. Mol. Opt. Phys. 26, 79 (1990).

[32] P. C. Stancil, B. Zygelman, K. Kirby, in *Photonic, Electronic, and Atomic Collisions*, edited by F. Aumayr and H. P. Winter (World Scientific, Singapore, 1998), p. 537.

[33] L. F. Errea, J. D. Gorfinkiel, A. Macías, L. Méndez and A. Riera, J. Phys. B: At. Mol. Opt. Phys. 30, 3855 (1997); L. F. Errea, A. Macías, L. Méndez, I. Rabadán and A. Riera, Int. J. Mol. Sci. 3, 142 (2002).

[34] V. Sidis, Adv. At. Mol. Opt. Phys. 26, 161 (1990).

[35] J. B. Delos, Rev. Mod. Phys. 53, 287 (1981).

[36] L. F. Errea, L. Méndez, A. Riera, J. Phys. B 15, 101 (1982).

[37] F. Fraija, A. R. Allouche, M.-C. Bacchus-Montabonel, Phys. Rev. A 49, 272 (1994).

[38] MOLPRO package of *ab initio* programs designed by H. J. Werner and P. Knowles, version 2010.1. See http://www.molpro.net

[39] T. H. Dunning, Jr., J. Chem. Phys. 90, 1007 (1989).

[40] E. Baloïtcha, M. Desouter-Lecomte, M.-C. Bacchus-Montabonel, and N. Vaeck, J. Chem. Phys. 114, 8741 (2001).

[41] M.-C. Bacchus-Montabonel, Phys. Rev. A 46, 217 (1992).

[42] M.-C. Bacchus-Montabonel, F. Fraija, Phys. Rev. A 49, 5108 (1994).

[43] NIST Atomic Spectra Database Levels Data, [http://physics.nist.gov/cgi-bin/AtData/main_asd]

[44] K. P. Huber and G. Herzberg, in *Molecular Spectra and Molecular Structure IV. Constants of Diatomic Molecules* (Van Nostrand Reinhold, New York, 1979).

[45] R. J. Allan, C. Courbin, P. Salas, P. Wahnon, J. Phys. B 23, L461 (1990).

[46] D.R. Bates, R. McCarroll, Proc. R. Soc. A 245, 175 (1958).

[47] T.G. Winter, G.J. Hatton, Phys. Rev. A 21, 793 (1980).

[48] P. Honvault, M.-C. Bacchus-Montabonel, and R. McCarroll, J. Phys. B 27, 3115 (1994).

[49] M.C. Bacchus-Montabonel, Phys. Rev. A **40**, 6088 (1989).

**Table 1.** Comparison of asymptotic energies (in a.u.) from separated species at the optimized $NO(^2\Pi)$ distance ρ=2.187 a.u. in the case of singlet spin multiplicity.

| Configuration | CASSCF calculation for $HeNO^+$ at R=12 a.u. Linear geometry | Exp+CASSCF calculation for separated species | Exp+MRCI calculation for separated species |
|---|---|---|---|
| $He + NO^+(2^1\Delta)$ | 0.652 | 0.669 | 0.631 |
| $He + NO^+(3^1\Pi)$ | 0.635 | 0.645 | 0.613 |
| $He^+ + NO(^2\Pi)$ | 0.587 | 0.608 | 0.559 |
| $He + NO^+(2^1\Sigma^+)$ | 0.551 | 0.562 | 0.530 |
| $He + NO^+(2^1\Pi)$ | 0.520 | 0.529 | 0.502 |
| $He + NO^+(1^1\Sigma^-)$ | 0.346 | 0.355 | 0.339 |
| $He + NO^+(1^1\Delta)$ | 0.337 | 0.346 | 0.325 |
| $He + NO^+(1^1\Pi)$ | 0.332 | 0.336 | 0.321 |
| $He + NO^+(1^1\Sigma^+)$ | 0.0 | 0.0 | 0.0 |

**Table 2**. Comparison of asymptotic energies (in a.u.) from separated species at the optimized $NO(^2\Pi)$ distance ρ=2.187 a.u. in the case of triplet spin multiplicity.

| Configuration | CASSCF calculation for $HeNO^+$ at R=12 a.u. Linear geometry | Exp+CASSCF calculation for separated species | Exp+MRCI calculation for separated species |
|---|---|---|---|
| $He + NO^+(2^3\Sigma^-)$ | 0.362 | 0.359 | 0.347 |
| $He + NO^+(4^3\Pi)$ | 0.354 | 0.348 | 0.335 |
| $He + NO^+(2^3\Delta)$ | 0.352 | 0.349 | 0.342 |
| $He^+ + NO(^2\Pi)$ | 0.347 | 0.355 | 0.318 |
| $He + NO^+(3^3\Pi)$ | 0.293 | 0.290 | 0.285 |
| $He + NO^+(2^3\Pi)$ | 0.208 | 0.211 | 0.204 |
| $He + NO^+(1^3\Sigma^-)$ | 0.075 | 0.070 | 0.074 |
| $He + NO^+(1^3\Delta)$ | 0.053 | 0.056 | 0.047 |
| $He + NO^+(1^3\Pi)$ | 0.012 | 0.008 | 0.012 |
| $He + NO^+(1^3\Sigma^+)$ | 0.0 | 0.0 | 0.0 |

**Table 3**: Total and partial charge transfer cross sections for the $He^+$ + NO collision system (perpendicular geometry, θ=90°) (in $10^{-16}$ $cm^2$).

| | | | $He^+$ + NO singlet states | | | $He^+$ + NO triplet states | | | S+T |
|---|---|---|---|---|---|---|---|---|---|
| v(au) | $E_{CM}$ (keV) | Elab (keV) | σ A'states | σ A''states | $σtot^T$ | σ A'states | σ A''states | $σtot^S$ | $σtot^{S+T}$ |
| 0.07 | 0.43 | 0.49 | 11.10 | 0.04 | 11.14 | 12.98 | 0.24 | 13.22 | 12.70 |
| 0.1 | 0.88 | 1. | 11.57 | 0.07 | 11.64 | 11.17 | 0.30 | 11.14 | 11.26 |
| 0.12 | 1.27 | 1.44 | 10.09 | 0.09 | 10.18 | 9.30 | 0.33 | 9.63 | 9.77 |
| 0.14 | 1.73 | 1.96 | 9.81 | 0.10 | 9.91 | 8.94 | 0.31 | 9.25 | 9.41 |
| 0.15 | 1.98 | 2.25 | 9.17 | 0.09 | 9.26 | 8.49 | 0.28 | 8.77 | 8.89 |
| 0.17 | 2.54 | 2.89 | 8.84 | 0.09 | 8.93 | 8.03 | 0.28 | 8.31 | 8.46 |
| 0.18 | 2.86 | 3.24 | 8.19 | 0.01 | 8.28 | 6.90 | 0.27 | 7.18 | 7.45 |
| 0.2 | 3.53 | 4. | 6.81 | 0.11 | 6.92 | 5.79 | 0.33 | 6.12 | 6.32 |
| 0.25 | 5.51 | 6.25 | 5.73 | 0.17 | 5.90 | 5.30 | 0.41 | 5.71 | 5.76 |
| 0.3 | 7.94 | 9. | 4.36 | 0.34 | 4.70 | 3.83 | 0.49 | 4.32 | 4.41 |
| 0.4 | 14.11 | 16. | 3.60 | 0.44 | 4.05 | 3.11 | 0.60 | 3.71 | 3.79 |
| 0.5 | 22.05 | 25. | 3.07 | 0.50 | 3.57 | 2.25 | 0.73 | 2.98 | 3.13 |

## Figure caption

Fig. 1: Internal Jacobi coordinates for the $HeNO^+$ molecular system.

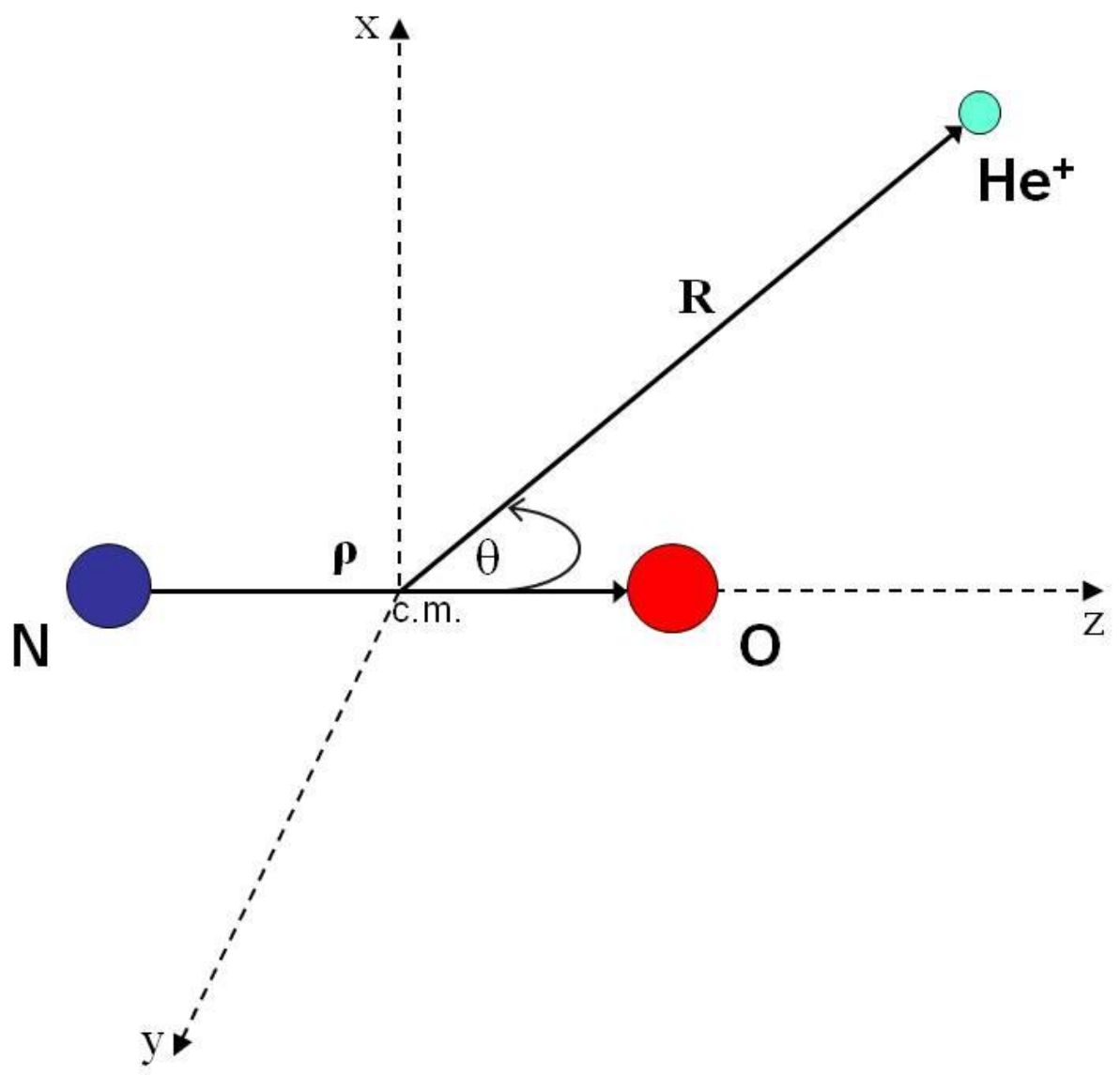

Fig. 2: Potential energy curves of the $(He\text{-}NO)^+$ molecular system (linear geometry, $\theta$=180°). (a) singlet states; (b) triplet states. —— Π states; —— $\Sigma^+$ states; ---- Δ states; —— $\Sigma^-$ states.

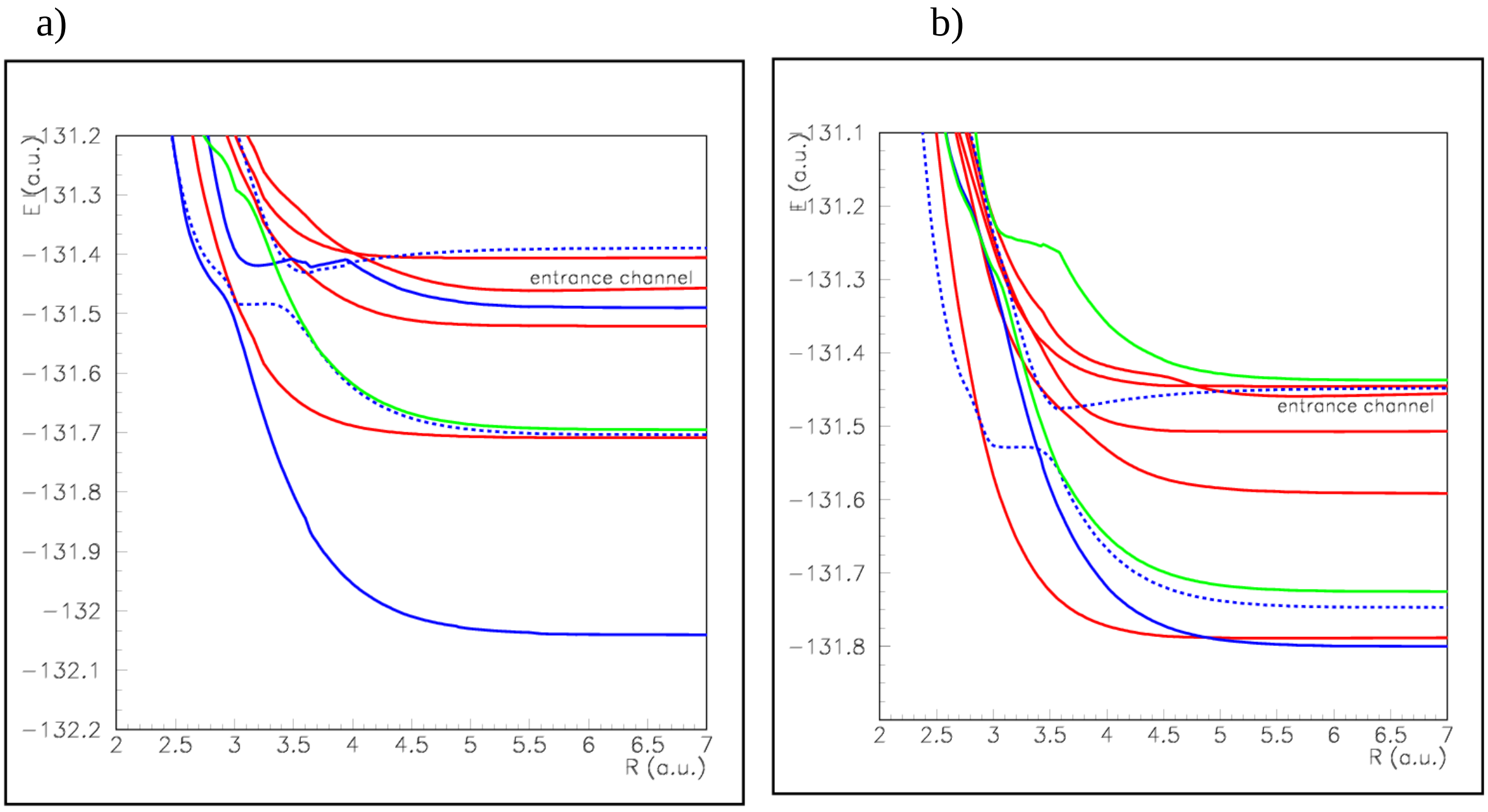

Fig. 3: Potential energy curves of the singlet $(He\text{-}ON)^+$ molecular states (linear geometry, $\theta$=0°). —— $\Pi$ states; —— $\Sigma^+$ states; ---- $\Delta$ states; —— $\Sigma^-$ states.

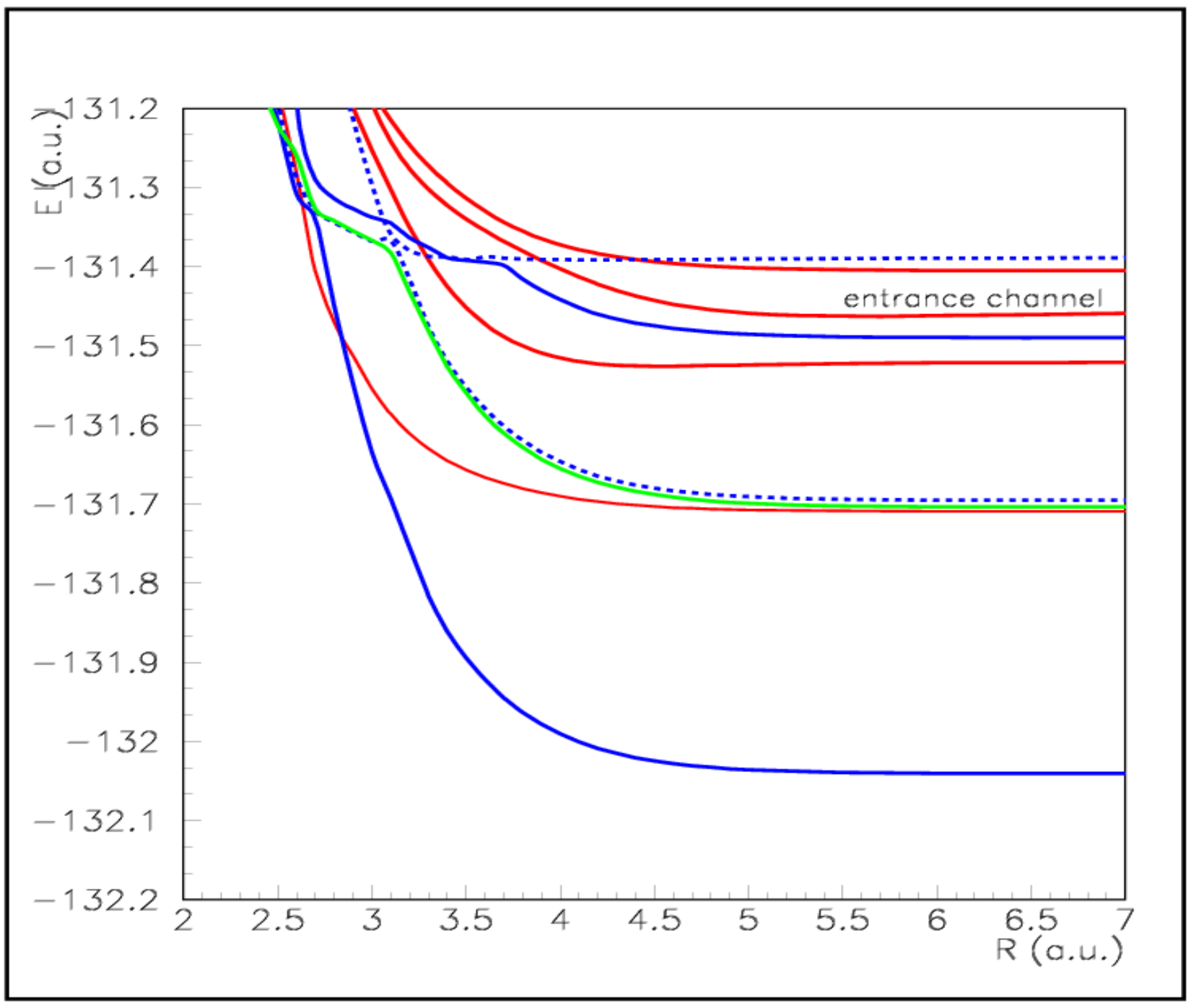

Fig. 4: Potential energy curves of the singlet states of the $(He\text{-}NO)^+$ molecular system.

(a) perpendicular geometry, $\theta$=90°; (b) $\theta$=45°. —— A' states; ---- A'' states.

(a) (b)

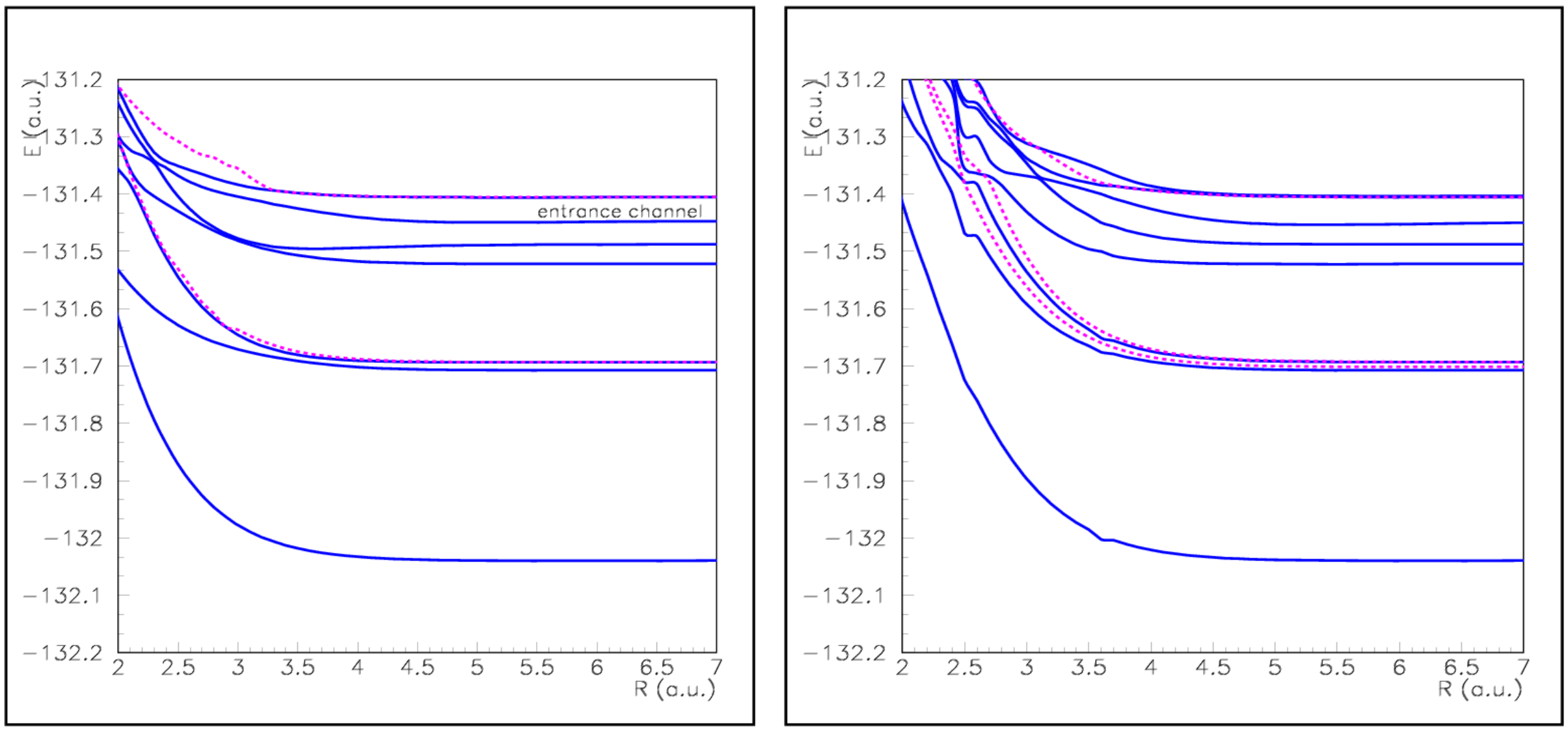

Fig. 5: Comparison of total cross sections in the perpendicular approach with regard to the experimental data (in $10^{-16}$ cm$^2$).

— triplet manifold ; — singlet manifold; — mean value; — experiment [15]

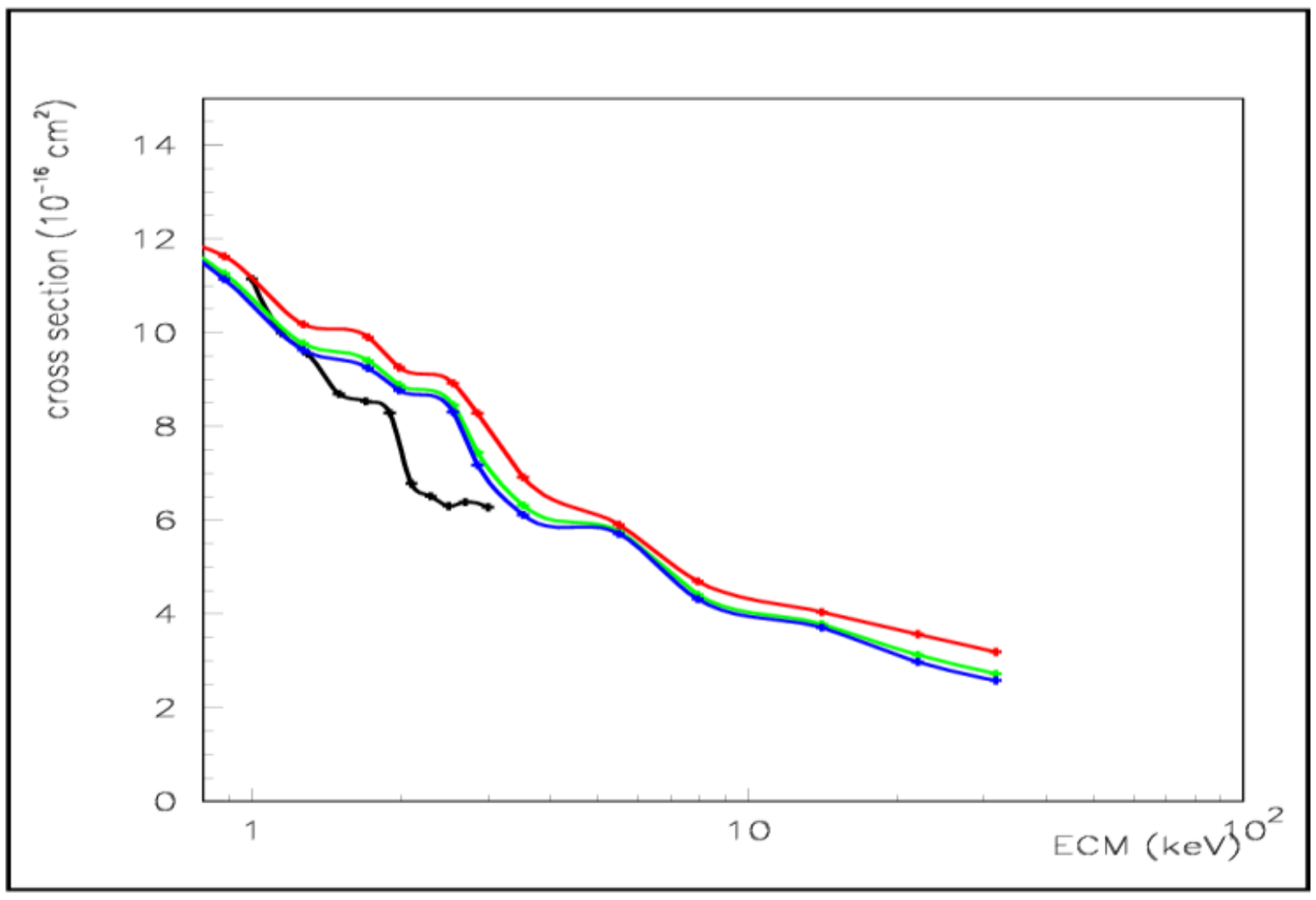

Fig. 6: Comparison of total cross sections with regard to the orientation angle (in $10^{-16}$ cm$^2$).

— θ=180°; — θ= 0°; — θ= 45°; — θ= 90°; — experiment [15]

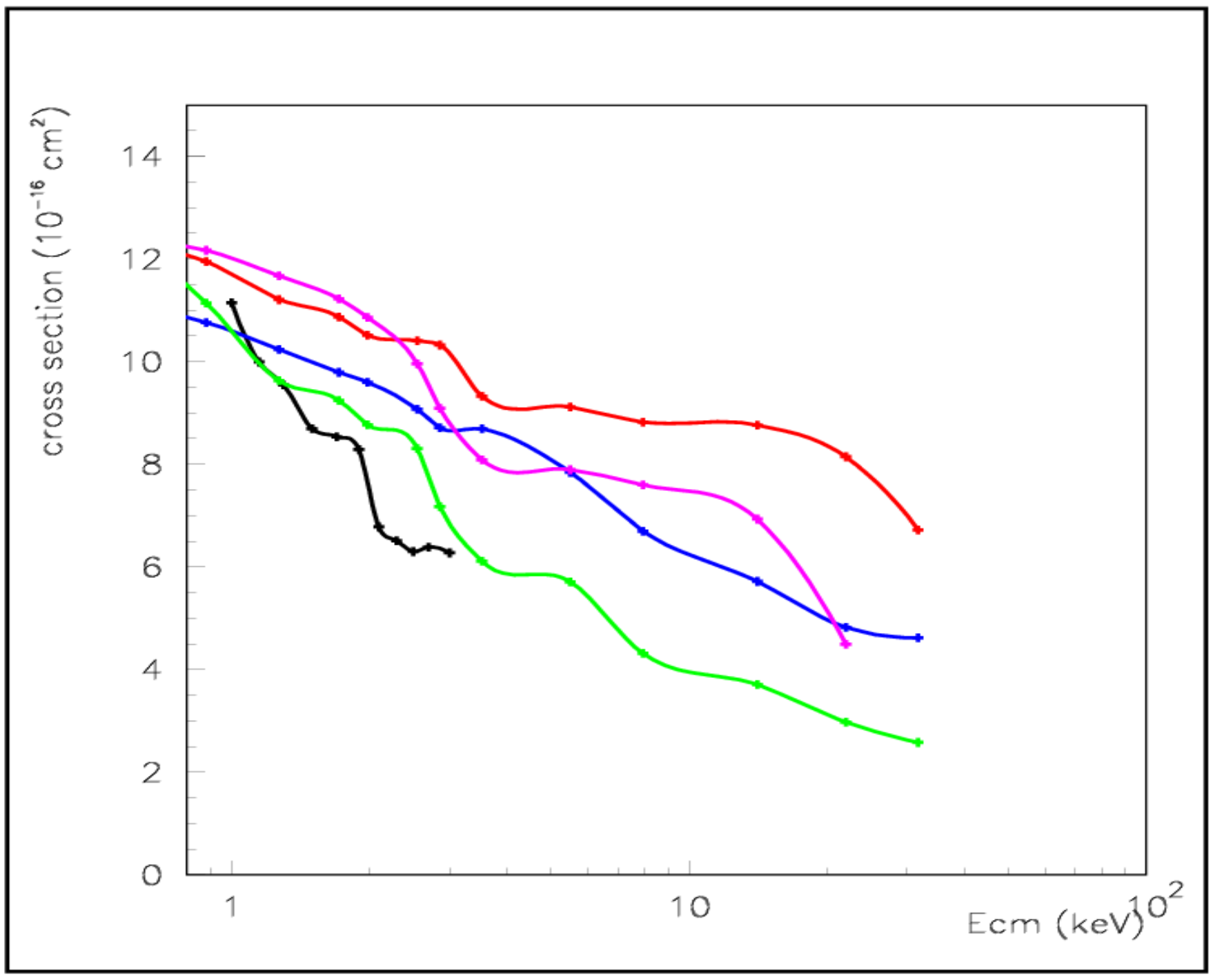